\documentclass[a4paper,twoside]{jpconf-vaso}
\usepackage{graphicx}
\usepackage{hyperref,cite,xspace,amsmath,amssymb}
\usepackage{url}
\newcommand{\none}{\ensuremath{\tilde{\chi}_1^0}\xspace}
\newcommand{\ntwo}{\ensuremath{\tilde{\chi}_2^0}\xspace}
\newcommand{\chone}{\ensuremath{\tilde{\chi}_1^{\pm}}\xspace}
\newcommand{\glu}{\ensuremath{\tilde{g}}\xspace}
\newcommand{\sq}{\ensuremath{\tilde{q}}\xspace}
\newcommand{\st}{\ensuremath{\tilde{t}}\xspace}
\newcommand{\sbb}{\ensuremath{\tilde{b}}\xspace}
\newcommand{\stau}{\ensuremath{\tilde{\tau}_1}\xspace}
\newcommand{\staur}{\ensuremath{\tilde{\tau}_{\mathrm R}}\xspace}
\newcommand{\gra}{\ensuremath{\tilde{G}}\xspace}
\newcommand{\piplus}{\ensuremath{\pi^{\pm}}\xspace}
\newcommand{\ifb}{\mbox{fb$^{-1}$}\xspace}

\newcommand{\tev}{\mbox{TeV}\xspace}
\newcommand{\gev}{\mbox{GeV}\xspace}

\newcommand{\madnlo}{\textsc{MadGraph5}\_a\textsc{MC@NLO}\xspace}
\newcommand{\pyt}{\textsc{Pythia8}\xspace}

\DeclareGraphicsExtensions{.pdf,.png,.jpg}
\graphicspath{{./}{./figs/}}

\begin{document}

\markright{}
\begin{flushright}
IFIC/19-16 \\
KCL-PH-TH/2019-34
\end{flushright}
\vspace*{-0.78cm}

\title{SUSY discovery prospects with MoEDAL}

\author{K Sakurai$^1$, D Felea$^2$, J Mamuzic$^3$, N E Mavromatos$^4$, V A Mitsou$^{3,\ast}$, J L Pinfold$^5$, R Ruiz de Austri$^3$, A Santra$^3$ and O Vives$^3$}

\address{$^1$ Institute of Theoretical Physics, Faculty of Physics, University of Warsaw, ul.\ Pasteura 5, PL02093 Warsaw, Poland}

\address{$^2$ Institute of Space Science, P.O.\ Box MG~23, 077125, Bucharest -- M\u{a}gurele, Romania}

\address{$^3$ Instituto de F\'isica Corpuscular (IFIC), CSIC -- Universitat de Val\`encia, C/ Catedr\'atico Jos\'e Beltr\'an 2, E-46980 Paterna (Valencia), Spain}

\address{$^4$ Theoretical Particle Physics and Cosmology Group, Department of Physics, King's College London, Strand, London WC2R 2LS, UK}

\address{$^5$ Physics Department, University of Alberta, Edmonton Alberta T6G~2E4, Canada}

\ead{vasiliki.mitsou@ific.uv.es}

\begin{abstract}
We present a preliminary study on the possibility to search for massive long-lived electrically charged particles at the MoEDAL detector. MoEDAL is sensitive to highly ionising objects such as magnetic monopoles or massive (meta-)stable electrically charged particles and we focus on the latter in this paper. Requirements on triggering or reducing the cosmic-ray and cavern background, applied in the ATLAS and CMS analyses for long-lived particles, are not necessary at MoEDAL, due to its completely different detector design and extremely low background. On the other hand, MoEDAL requires slow-moving particles, resulting in sensitivity to massive states with typically small production cross sections. Using Monte Carlo simulations, we compare the sensitivities of MoEDAL versus ATLAS/CMS for various long-lived particles in supersymmetric models, and we seek a scenario where MoEDAL can provide discovery reach complementary to ATLAS and CMS.
\end{abstract}

\markright{K Sakurai \textit{et al.}}

\section{Introduction}\label{sc:intro}

Supersymmetry (SUSY) is an extension of the Standard Model (SM) which assigns to each SM field a superpartner field with a spin differing by a half unit. SUSY provides elegant solutions to several open issues in the SM, such as the hierarchy problem, the identity of dark matter, and the grand unification. In some SUSY scenarios, the existence of long-lived particles (LLPs) is predicted~\cite{Fairbairn:2006gg}, that may either decay within the typical volume of an LHC detector or may traverse it entirely as (meta-)stable. In the former case, it may give rise to displaced vertices~\cite{Aaboud:2018aqj,Sirunyan:2018vlw} or disappearing tracks~\cite{Aaboud:2017mpt,Sirunyan:2018ldc}. Here we focus on the latter case and more precisely on heavy, stable charged particles (HSCPs)\footnote{If the stable particle is neutral, hence only weakly interacting, such as the \none, its signature is the typical for SUSY searches of large missing transverse momentum and therefore it is not discussed in the context of LLPs.} that may give rise to anomalous ionisation. 

HSCPs may be observed in experiments sensitive to high ionisation, both in collider experiments~\cite{Lee:2018pag,Alimena:2019zri} and in cosmic observatories~\cite{Burdin:2014xma}.  In particular, the general-purpose ATLAS and CMS experiments at the Large Hadron Collider (LHC) have searched for and set limits in such scenarios. Besides them, dedicated detectors are being proposed to explore this less-constrained manifestations of physics beyond the SM~\cite{Alimena:2019zri}.  Among them, the Monopole and Exotics Detector At the LHC (MoEDAL)~\cite{Pinfold:2009oia} is the only one in operation as of today. It is specially designed to detect highly ionising particles (HIPs) such as magnetic monopoles and HSCPs, covering a wide spectrum of theoretical proposals~\cite{Acharya:2014nyr}, in a manner complementary to ATLAS and CMS~\cite{DeRoeck:2011aa}. 

The structure of this paper is as follows. In section~\ref{sc:hscp}, SUSY models predicting HSCPs are discussed, reviewing also their current experimental bounds. An overview of the MoEDAL detecting systems and analysis techniques, emphasising the complementarity to the experimental approach followed in ATLAS/CMS is given in section~\ref{sc:moedal}. The SUSY HSCP kinematics in MoEDAL are studied in section~\ref{sc:direct}. Section~\ref{sc:realistic} presents some preliminary results from a case study of simplified topologies where MoEDAL can be sensitive to regions of the parameter space different than ATLAS/CMS. We close this report with some concluding remarks and an outlook in section~\ref{sc:concl}.

\section{HSCPs \& SUSY @ LHC}\label{sc:hscp}

In supersymmetric models, various instances of sparticles/objects may emerge as HSCPs. Considering its detector placements in the cavern and its low-velocity sensitivity, MoEDAL may detect HSCPs with proper lifetimes $c\tau\gtrsim1$~m. 

\begin{description}

\item[Sleptons:] They may be long-lived as next-to-the-lightest SUSY partners (NLSPs) decaying to a gravitino or a neutralino LSP. In gauge-mediated symmetry breaking (GMSB), the \stau NLSP decays to \gra may be suppressed due to the ``weak'' gravitational interaction~\cite{Hamaguchi:2006vu}, remaining partly compatible with constraints on the dark-matter abundance in super-weakly interacting massive particle scenarios~\cite{Feng:2015wqa}. In other cases, such as the  co-annihilation region in constrained MSSM, the most natural candidate for the NLSP is the lighter \stau, which could be long lived if the mass splitting between the \stau and the \none is small~\cite{Jittoh:2005pq,Kaneko:2008re,Feng:2015wqa}.
This region is one of the most favoured by the measured dark-matter relic density~\cite{Ellis:2003cw}.

\item[R-hadrons:] They are formed by hadronised metastable gluinos, stops or sbottoms. Gluino R-hadrons arise in Split SUSY due to the ultra-heavy squarks strongly suppressing \glu decays to \sq and quarks~\cite{ArkaniHamed:2004fb,ArkaniHamed:2004yi}. Other models, such as $R$-parity-violating SUSY~\cite{Evans:2012bf} or gravitino dark matter~\cite{DiazCruz:2007fc}, could produce a long-lived squark (\st or \sbb)  that would also form an R-hadron. 

\item[Charginos:] Their long lifetime may be due to their mass degeneracy with the \none LSP in anomaly-mediated symmetry breaking (AMSB) scenarios~\cite{Giudice:1998xp,Randall:1998uk}. When they decay within the detectors to a soft \piplus and a \none, they manifest themselves as disappearing tracks~\cite{Aaboud:2017mpt,Sirunyan:2018ldc}. 

\end{description}
 
ATLAS and CMS have searched for stable sleptons, R-hadrons and charginos using anomalously high energy deposits in the silicon tracker and long time-of-flight measurements by the muon system. The (very recent) ATLAS analysis~\cite{Aaboud:2019trc} has set the most stringent limits with 31.6~\ifb of $pp$ collisions at 13~\tev, while the CMS has used 2.5~\ifb so far~\cite{Khachatryan:2016sfv}.  The ATLAS bounds at 95\% confidence limit (CL) are 2000~\gev for gluino R-hadrons, 1250~\gev for sbottom R-hadrons, 1340~\gev for stop R-hadrons,  430~\gev for sleptons and 1090~\gev for charginos with sufficiently long lifetime~\cite{Aaboud:2019trc}. In refs.~\cite{ATLASsummary,CMSsummary}, summary plots of ATLAS and CMS analyses results pertaining to HSCPs are provided. For comprehensive and recent reviews on LHC past, current and future searches, the reader is referred to refs.~\cite{Lee:2018pag,Alimena:2019zri}. 

\section{MoEDAL and complementarity to ATLAS and CMS}\label{sc:moedal}

The MoEDAL detector~\cite{Pinfold:2009oia} is deployed around the intersection region at LHC Point~8 (IP8) in the LHCb vertex locator cavern.  It is a unique and largely passive detector comprising different detector technologies.

The MoEDAL main sub-detectors are made of a large array of CR-39, Makrofol\textregistered\ and Lexan\texttrademark\ nuclear track detector (NTD) stacks surrounding the intersection area. The passage of a HIP through the plastic sheet is marked by an invisible damage zone along the trajectory, which is revealed as a cone-shaped etch-pit when the plastic detector is chemically etched. Then the detector is scanned looking for aligned etch pits in multiple sheets. The MoEDAL NTDs have a threshold of $z/\beta\sim5$, where $z$ is the charge and $\beta=v/c$ the velocity of the incident particle.

A unique feature of the MoEDAL detector is the use of paramagnetic magnetic-monopole trappers (MMTs) to capture charged HIPs. In monopoles, the high magnetic charge implies a strong magnetic dipole moment, which may result in strong binding of the monopole with the nuclei of the aluminium MMTs. In such a case, the presence of a trapped monopole would be detected in a superconducting magnetometer through the induction technique~\cite{Acharya:2019vtb}. 

The MMTs may also capture HSCPs, which can only be observed through the detection of their decaying products. To this effect, the MoEDAL Collaboration is planning the MoEDAL Apparatus for detecting extremely Long Lived particles (MALL)~\cite{Alimena:2019zri}. In this case, MoEDAL MMTs, after they have been scanned through a magnetometer to identify any trapped monopole, will be shipped to a remote underground facility to be monitored for the decay of HSCPs. MALL is sensitive to charged particles and to photons, with energy as small as 1~GeV, and will monitor the MMTs for decays of captured particles. 

Given the unique design of the MoEDAL subsystems, the complementary aspects of MoEDAL to ATLAS/CMS, as far as HSCPs are concerned, come at no surprise.  MoEDAL is practically ``time-agnostic'' due to the passive nature of its detectors. Therefore, signal from very slowly moving particles will not be lost due to arriving in several consecutive bunch crossings. In addition, ATLAS/CMS carry out triggered-based analyses relying either on triggering on accompanying ``objects'', e.g.\ missing transverse momentum, or by developing and applying specialised triggers. 

MoEDAL, on the other hand, is mainly limited by the lower luminosity delivered at IP8, by the geometrical acceptance of the detectors, especially the MMTs, and by the requirement of passing the $z/\beta$ threshold of NTDs. In general, ATLAS and CMS have demonstrated their ability to cover high velocities, while MoEDAL is sensitive to lower ones $\beta \lesssim 0.2$. Typically $\beta \gtrsim 0.5$ is a safe limit for ATLAS/CMS, otherwise information moves to a different bunch crossing, making it very difficult to reconstruct, if at all possible.

Both ATLAS and CMS have to select the interesting events out of a huge background of known processes which may fake the sought-after events. To suppress this background, they have to apply offline cuts that unavoidably limit the efficiency of LLP detection, thus reducing the  parameter space probed by ATLAS/CMS. On the other hand, MoEDAL has practically no background and requires no trigger or selection cuts to detect a HIP, therefore it may detect particles that may escape detection at other LHC experiments. If the selection criteria imposed by ATLAS/CMS when searching for LLPs are left out in MoEDAL, parameter space otherwise uncovered can be explored by MoEDAL, as will be shown in section~\ref{sc:realistic}.

Regarding particles stopped in material and their subsequent decays, different approaches are followed. ATLAS/CMS look in empty bunch crossings for decays of trapped particles into jets~\cite{Aad:2013gva,Sirunyan:2017sbs}. The background in MALL, on the other hand, coming mainly from cosmic rays, should be easier to control and assess. The probed lifetimes --- up to $\sim\!10$~years according to initial estimates --- should be larger due to the unlimited monitoring time. 

\section{Direct production of metastable sparticles at LHC}\label{sc:direct}

In this study, we are discussing the kinematics of metastable sparticles in 13~\tev $pp$ collisions, focusing on their velocity $\beta$, which is the figure of merit for MoEDAL. We use \madnlo~\cite{Alwall:2014hca} and \pyt~\cite{Sjostrand:2007gs} to generate \staur pairs in direct production. The $\beta$ distributions for various \staur masses are shown in figure~\ref{fg:stau-direct}. The fraction of events with  $\beta \lesssim 0.2$, i.e.\ within the range of NTD sensitivity, only becomes significant for large \staur masses of ${\mathcal O}(1~\tev)$. In this mass range, the cross section is very low, making the possibility for \staur detection in the NTDs marginal.
 
\begin{figure}[tb]
\begin{minipage}{0.47\linewidth}
\includegraphics[width=\linewidth]{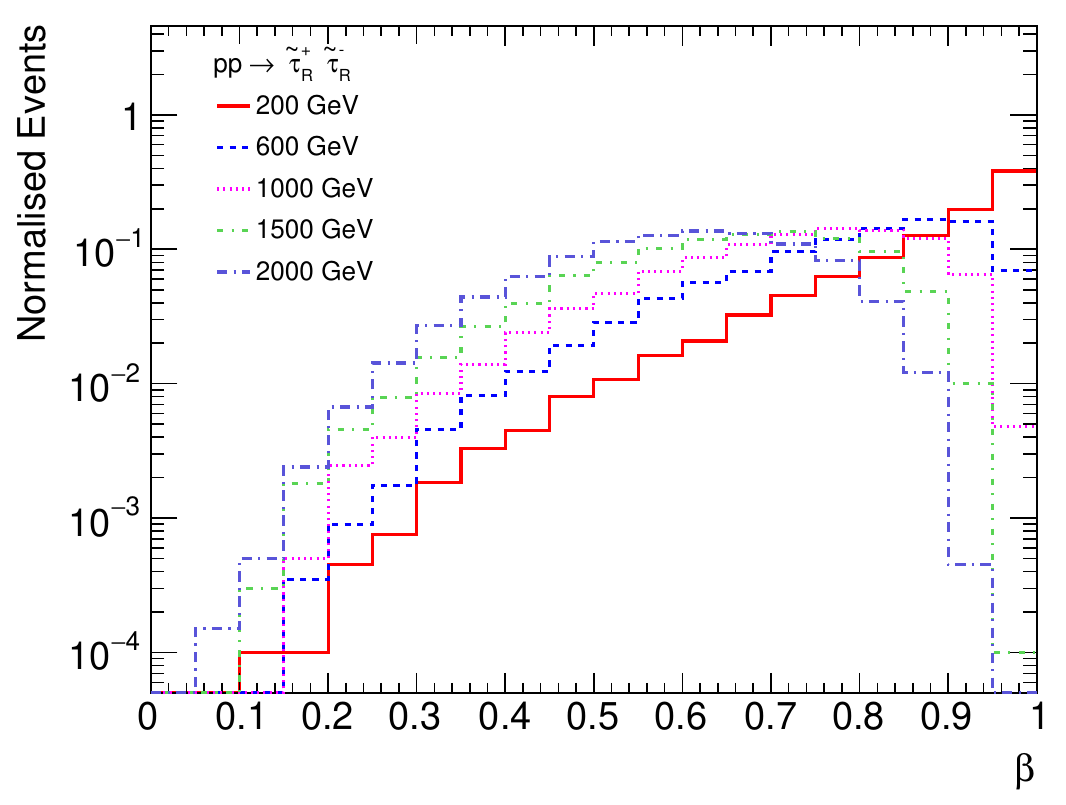}
\caption{\label{fg:stau-direct}Stau velocity distributions for $\tilde{\tau}_{\rm R}^+\tilde{\tau}_{\rm R}^-$ direct production in 13~\tev $pp$ collisions for  stau masses between 200~\gev and 2~\tev.}
\end{minipage}\hspace{0.05\linewidth}%
\begin{minipage}{0.47\linewidth}
\includegraphics[width=\linewidth]{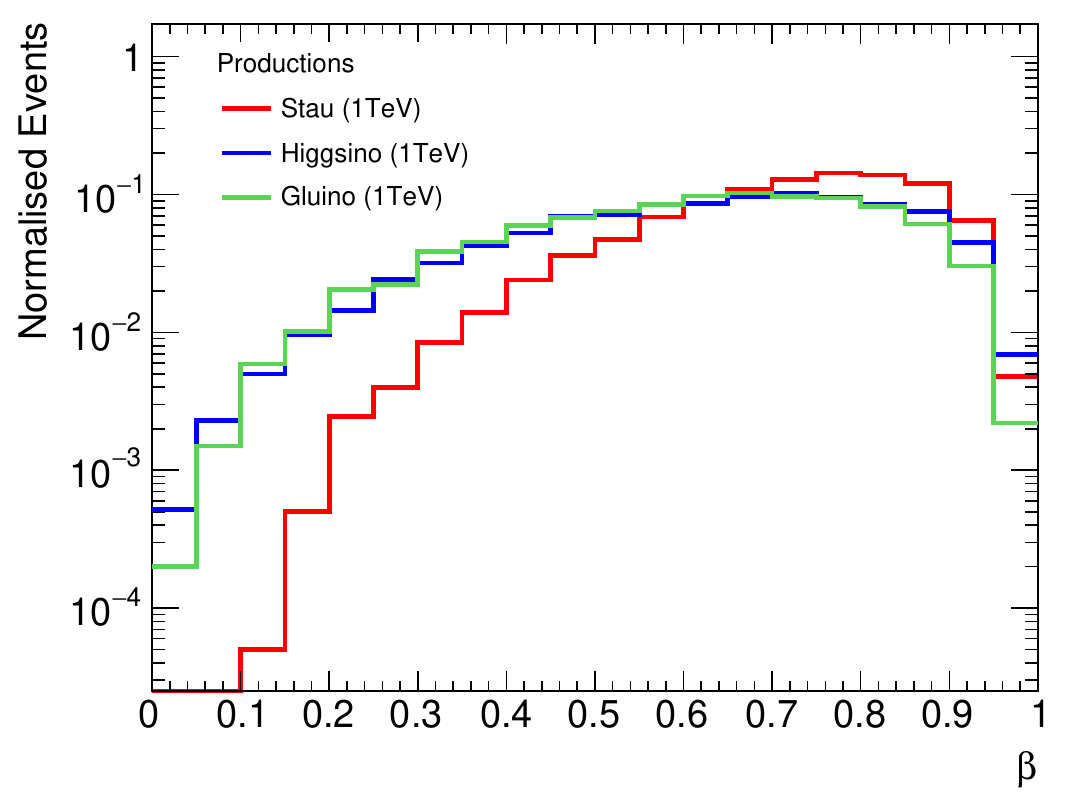}
\caption{\label{fg:fermion-boson}Comparison of velocity distributions between staus, higgsinos and gluinos of the same mass ($1~\tev$) pair produced directly in 13~\tev $pp$ collisions.}
\end{minipage} 
\end{figure}

Looking for alternatives, we have also simulated the direct pair production of higgsinos ($\none\,\chone$, $\ntwo\,\chone$) and gluinos ($\glu\,\!\glu$), besides that of staus ($\tilde{\tau}_{\rm R}^+\tilde{\tau}_{\rm R}^-$). As shown in their $\beta$ distributions in figure~\ref{fg:fermion-boson}, fermions (gluinos, hisggsinos) are slower than bosons (staus), hence more highly ionising. This is because the dominant channel is an $s$-channel spin-1 gauge boson $(Z^*/\gamma^*)$ exchange with $q\bar{q}$ initial states. The gauge bosons are transversely polarised due to helicity conservation in the initial vertex, so the final state must have a total nonzero angular momentum. The scalar (spinless) pair production (stau) undergoes a $p$-wave suppression, i.e.\ the production cross section vanishes as the stau velocity goes to zero to conserve angular momentum. No such suppression exists in the fermion (spinful) case. Between higgsinos and gluinos, the latter would be preferable in this context as they are produced more abundantly. To conclude, gluino pair direct production would serve as the best scenario for MoEDAL, since they are heavy fermions with large cross section. In the following, we discuss \stau as HSCP produced in \glu cascade decays, leaving the study of \glu R-hadrons for the future. 

\section{MoEDAL versus CMS: a preliminary case study}\label{sc:realistic}

As discussed earlier, we concentrate our efforts on heavy long-lived sparticles with a large production cross section that, in addition, satisfy present bounds. Therefore, we do not only study the MoEDAL sensitivity, but we also contrast it with ATLAS/CMS expected results. The latter is achieved by making use of the CMS efficiencies for HSPCs published in ref.~\cite{Khachatryan:2015lla}, which were extracted in order to reinterpret previous HSCP search results by CMS~\cite{Chatrchyan:2013oca} in specific supersymmetric models. 

To model the MoEDAL detector response, we assume an NTD efficiency $\varepsilon$ defined independently of the incident angle as
\begin{eqnarray}\label{eq:efficiency}
\varepsilon = 
\begin{cases}
1, \quad \beta \leq \beta_\text{max}, \\
0, \quad \beta > \beta_\text{max}, \qquad \text{where } \beta_\text{max} = 0.1 \sim 0.2.
\end{cases}
\end{eqnarray}

The geometrical coverage of the NTD is modelled by a 2-m radius sphere, although in reality some NTD panels are $<0.5$~m away from the IP. To account for the non-hermetic NTD geometry, we considered the NTD coverage fraction in the azimuth $\phi$ as a function of pseudorapidity $\eta$. For the 2015 NTD deployment, this amounts to a geometrical acceptance of 18.7\%.

Following the conclusions of section~\ref{sc:direct}, we assume \glu pair production, attempting to identify cascade decays of the \glu to a \stau which may evade the selection criteria applied in ref.~\cite{Chatrchyan:2013oca}, hence weakening the exclusion limits set by CMS, while maintaining high MoEDAL efficiency, i.e.\ slow-moving staus. In particular, we focus on two of the selection cuts:

\begin{description}

\item[At least one Pixel hit:] Requires the presence of a charged particle in the innermost part of the detector. The event may be rejected by the CMS analysis, if the \glu decays via a long-lived neutral particle, e.g.\ a neutralino.

\item[Small impact parameters $\boldsymbol{d_z, d_{xy} < 0.5}$~cm:] The long-lived charged track must point back to the primary vertex. It is imposed against cosmic-ray background. However, if a particle in the decay chain is long-lived and a kink is present, the event may be missed by CMS.

\end{description}

In the following, we require at least \emph{two hits\/} in the NTDs to deem a point in the parameter space as observable by MoEDAL, which represents a rather tight requirement, given the extremely low background expected. For CMS we only show 95\% CL exclusion regions, which are typically more extensive than the discovery ones. In this respect, the comparison is rather biased favouring CMS, yet it serves well in the context of a preliminary study.

Concerning the datasets, we show projections for two LHC runs:
\begin{itemize}
\item {\bf End of Run~2:} CMS has recorded $\sim150~\ifb$ of $pp$ collisions at $\sqrt{s}=13~\tev$ during 2015--2018, while MoEDAL was exposed to $\sim6.7~\ifb$ at IP8.\footnote{Due to the passive nature of the MoEDAL subsystems, the relevant integrated luminosity is the \emph{delivered\/} rather than the recorded as in other experiments.}
\item {\bf End of Run~3:} We assume a collision energy of $\sqrt{s}=14~\tev$ and 300~\ifb for ATLAS/CMS~\cite{Walkowiak:2018ief} to be collected during 2021--2023. The current scenario for LHCb is that it may receive roughly 10 times less luminosity than ATLAS/CMS in Run~3, a significant improvement over Run~2 where this figure was $\sim 50$. Hence, we assume 30~\ifb of delivered luminosity for MoEDAL by the end of 2023~\cite{LHCb:2018hiv}.
\end{itemize}

We generated the event samples with \madnlo and \pyt.  The event analyses for MoEDAL and CMS have been carried out at parton level.  In recasting the CMS analysis, we closely follow and use the recipe and efficiency maps provided in ref.~\cite{Khachatryan:2015lla}. Several decay chains were tried; here we highlight two cases. 

\subsection{Case~1: $\glu \to jj\none \to jj\piplus\stau$}

In this scenario, the \glu always decays to a \none, which decays to a \piplus and a \stau, provided that $m(\none)-m(\stau)<m(\tau)$. The small mass splitting between the neutralino and the stau makes the \none long lived, yet it decays in the inner detector if $m(\none)-m(\stau)=1~\gev$. The \none leaves no hit in the Pixel detector therefore the CMS efficiency suffers, while it does not affect the MoEDAL response. 

In figure~\ref{fg:N1-pi-stau}, we compare the MoEDAL discovery reach with the CMS exclusion limits when applying the analysis~\cite{Chatrchyan:2013oca} to the scenario of Case~1. The projected MoEDAL reach for Run~3 is barely better than the CMS expected, yet it clearly shows a different trend: MoEDAL may cover larger \none lifetimes, while it is weaker on the \glu mass mostly due the large luminosity needed to overcome the heavier, hence less abundant, gluinos. 
The principal reason for the poor improvement provided by MoEDAL compared to CMS is the reduced integrated luminosity; a factor of $\sim20$ less for Run~2 and $\sim10$ for Run~3. 

\begin{figure}[ht]
\includegraphics[width=0.58\linewidth]{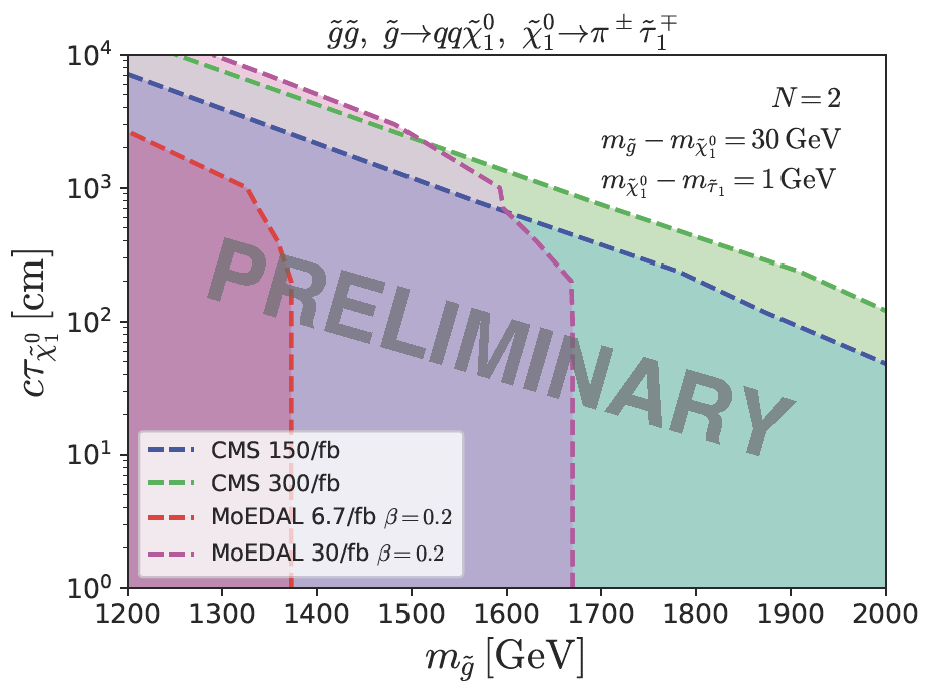}\hspace{0.05\linewidth}%
\begin{minipage}[b]{0.37\linewidth}\caption{\label{fg:N1-pi-stau}MoEDAL discovery reach requiring two signal events versus CMS 95\% CL exclusion plot in the \glu mass vs.\ \none lifetime plane for Run~2 (13~\tev) and for Run~3 (14~\tev) integrated luminosities. A scenario with gluino pair production where the \glu decays to a long-lived \none decaying to a metastable \stau and a \piplus is considered. The mass splitting between the \none and the \glu (\stau) is 30~\gev (1~\gev). A \stau detection threshold $z/\beta \geq 5 \Rightarrow \beta_{\text{max}} = 0.2$ is assumed.}
\end{minipage}
\end{figure}

\subsection{Case~2: $\glu \to jj\none \to jj\tau\stau$}

Here we make a further step, also attempting to evade the impact-parameter cuts. We assume that the \none decays to a \stau and a (heavy) $\tau$ instead of a (light) \piplus. The mass splitting between the \none and the \stau is fairly large (300~\gev), so the \none large lifetime is due to dynamical rather than kinematical reasons. Possibilities include an axino \none with a small coupling, or some vector-like fermion, which must be the neutral component of an $SU(2)$ doublet $D'$ that enjoys $u_R^c QD'$ and $\tau_R^c LD'$ couplings.  

The effect is shown in the diagram of figure~\ref{fg:decay}, where the \stau is produced with a kink, thus non-pointing back to the IP, escaping detection by CMS.  The reach comparison, shown in figure~\ref{fg:N1-tau-stau}, demonstrates clearly the effect of CMS missing a second cut: MoEDAL can improve significantly the sensitivity at large lifetimes.

\begin{figure}[ht]
\includegraphics[width=0.45\linewidth]{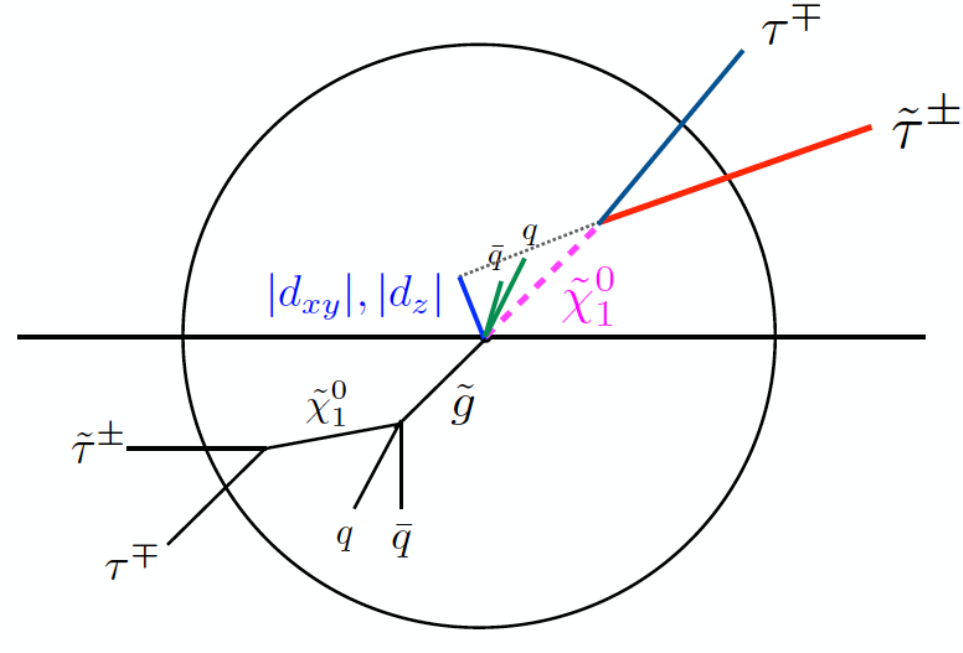}\hspace{0.05\linewidth}%
\begin{minipage}[b]{0.5\linewidth}\caption{\label{fg:decay} Gluino production and cascade decay considered in case~2. The \glu decays promptly while the \none is long-lived yet still decaying within the LHC detectors. The (heavy) tau in the \none decay gives rise to a kink that makes the \stau non-pointing.}
\end{minipage}
\end{figure}

\begin{figure}[ht]
\includegraphics[width=0.58\linewidth]{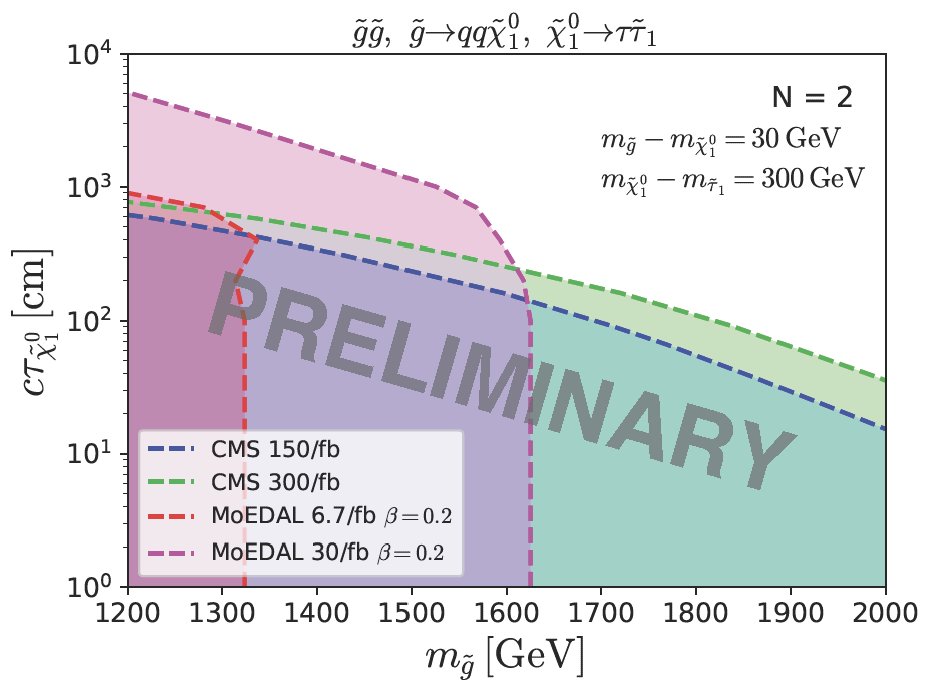}\hspace{0.05\linewidth}%
\begin{minipage}[b]{0.37\linewidth}\caption{\label{fg:N1-tau-stau}MoEDAL discovery reach requiring two signal events versus CMS 95\% CL exclusion plot in the \glu mass vs.\ \none lifetime plane for Run~2 (13~\tev) and for Run~3 (14~\tev) integrated luminosities. A scenario with gluino pair production where the \glu decays to a long-lived \none decaying to a metastable \stau and a $\tau$ is considered. The mass splitting between the \none and the \glu (\stau) is 30~\gev (300~\gev). A \stau detection threshold $z/\beta \geq 5 \Rightarrow \beta_{\text{max}} = 0.2$ is assumed.}
\end{minipage}
\end{figure}

As mentioned earlier, we consider a nominal value of $\beta_\text{max}=0.2$ for the stau velocity to yield an NTD hit. This value represents the best case for a CR-39 threshold of $z/\beta=5$, however larger threshold values may be necessary to suppress background during the NTD treatment. The MoEDAL sensitivity was assessed with alternative $\beta_\text{max}$ values of 0.15 ($z/\beta=6.7$) and 0.1 ($z/\beta=10$). It was shown that for a $z/\beta=6.7$, MoEDAL could still complement ATLAS/CMS in large \none lifetimes.

A word of caution is due here before drawing conclusions from this preliminary study: 
\begin{enumerate}
\item We only considered one of the CMS analysis (similar for ATLAS), however other analyses sensitive to HSCPs might cover part or the whole of the region extended by MoEDAL, such as ones targeting displaced vertices~\cite{Aaboud:2018aqj,Sirunyan:2018vlw} or disappearing tracks~\cite{Aaboud:2017mpt,Sirunyan:2018ldc}. Nonetheless, even in this case, the added value of MoEDAL would remain, since it provides a coverage with a completely different  detector and analysis technique, thus with uncorrelated systematic uncertainties.
\item The ATLAS/CMS selection cuts applied here were optimised for~7 and 8~\tev data, which are not expected to be optimal for~13 and 14~\tev collisions. Besides that, further past (Run~2) and future (Run~3) improvements in the analysis have been or will be made, that have not been taken into account here.
\item In the MoEDAL analysis, we did not take into account the incident angle of the sleptons to the NTDs nor the presence of material in front of the NTDs. 
\end{enumerate}

\section{Concluding remarks and outlook}\label{sc:concl}

We presented a preliminary study on the feasibility to detect massive metastable supersymmetric partners with the MoEDAL experiment in a complementary way to CMS, also valid for ATLAS. Direct production of heavy (hence slow-moving), fermions with large cross section (thus via strong interactions) is the most favourable scenario. MoEDAL is mostly sensitive to slow moving particles ($\beta \lesssim 0.2$) unlike ATLAS/CMS suitability for faster ones, yet the less integrated luminosity it receives at IP8 seems to be a limiting factor for simple scenarios, although these first results appear to be promising for more complex topologies.

More effort is needed towards several directions: (i) a realistic MoEDAL performance description; (ii) the exploration of realistic SUSY scenarios where the studied simplified topologies occur naturally; and (iii) an exhaustive study of ATLAS and CMS analyses sensitive to scenarios covered by MoEDAL. So far, we have only considered sleptons as the metastable particles that interact directly with the MoEDAL detectors; R-hadrons, and possibly charginos, also pose other possibilities worth examining in the future.  

\ack
We would like to thank colleagues from the MoEDAL Collaboration for discussions and comments and in particular J.~R.~Ellis and A.~de~Roeck. The work of KS is partially supported by the National Science Centre, Poland, under research grant 2017/26/E/ST2/00135 and the Beethoven grant DEC-2016/23/G/ST2/04301. The work of JM, VAM, RRA, AS and OV is supported by the Generalitat Valenciana via a special grant for MoEDAL and via the Project PROMETEO-II/2017/033; by the Spanish MICIU / AEI and the European Union / FEDER via the grants FPA2015-65652-C4-1-R, FPA2017-85985-P, FPA2017-84543-P and PGC2018-094856-B-I00; by the Severo Ochoa Excellence Centre Project SEV-2014-0398; and by a 2017 Leonardo Grant for Researchers and Cultural Creators, BBVA Foundation. The work of NEM is supported in part by the UK Science and Technology Facilities research Council (STFC) under the research grant ST/P000258/1 and by the Physics Department of King's College London. NEM also acknowledges a scientific associateship as \emph{Doctor Vinculado} at IFIC-CSIC-Valencia University, Spain. JLP acknowledges support by the Natural Science and Engineering Research Council of Canada via a project grant, by the V-P Research of the University of Alberta and by the Provost of the University of Alberta.

\section*{References}
\bibliography{Mitsou_Discrete2018}

\providecommand{\newblock}{}
\begin{thebibliography}{10}
\expandafter\ifx\csname url\endcsname\relax
  \def\url#1{{\tt #1}}\fi
\expandafter\ifx\csname urlprefix\endcsname\relax\def\urlprefix{URL }\fi
\providecommand{\eprint}[2][]{\url{#2}}

\bibitem{Fairbairn:2006gg}
Fairbairn M, Kraan A~C, Milstead D~A, Sjostrand T, Skands P~Z and Sloan T 2007
  {\em Phys. Rept.\/} {\bf 438} 1--63 (\textit{Preprint}
  \eprint{hep-ph/0611040})

\bibitem{Aaboud:2018aqj}
Aaboud M {\em et~al.\/} (ATLAS) 2019 {\em Phys. Rev.\/} {\bf D99} 052005
  (\textit{Preprint} \eprint{1811.07370})

\bibitem{Sirunyan:2018vlw}
Sirunyan A~M {\em et~al.\/} (CMS) 2019 {\em Phys. Rev.\/} {\bf D99} 032011
  (\textit{Preprint} \eprint{1811.07991})

\bibitem{Aaboud:2017mpt}
Aaboud M {\em et~al.\/} (ATLAS) 2018 {\em JHEP\/} {\bf 06} 022
  (\textit{Preprint} \eprint{1712.02118})

\bibitem{Sirunyan:2018ldc}
Sirunyan A~M {\em et~al.\/} (CMS) 2018 {\em JHEP\/} {\bf 08} 016
  (\textit{Preprint} \eprint{1804.07321})

\bibitem{Lee:2018pag}
Lee L, Ohm C, Soffer A and Yu T~T 2019 {\em Prog. Part. Nucl. Phys.\/} {\bf
  106} 210--255 (\textit{Preprint} \eprint{1810.12602})

\bibitem{Alimena:2019zri}
Alimena J {\em et~al.\/} 2020 {\em J. Phys. G\/} {\bf 47} 090501
  (\textit{Preprint} \eprint{1903.04497})

\bibitem{Burdin:2014xma}
Burdin S, Fairbairn M, Mermod P, Milstead D, Pinfold J, Sloan T and Taylor W
  2015 {\em Phys. Rept.\/} {\bf 582} 1--52 (\textit{Preprint}
  \eprint{1410.1374})

\bibitem{Pinfold:2009oia}
Pinfold J {\em et~al.\/} (MoEDAL) 2009 {Technical Design Report of the MoEDAL
  Experiment} {CERN-LHCC-2009-006}, MoEDAL-TDR-001

\bibitem{Acharya:2014nyr}
Acharya B {\em et~al.\/} (MoEDAL) 2014 {\em Int. J. Mod. Phys.\/} {\bf A29}
  1430050 (\textit{Preprint} \eprint{1405.7662})

\bibitem{DeRoeck:2011aa}
De~Roeck A, Katre A, Mermod P, Milstead D and Sloan T 2012 {\em Eur. Phys.
  J.\/} {\bf C72} 1985 (\textit{Preprint} \eprint{1112.2999})

\bibitem{Hamaguchi:2006vu}
Hamaguchi K, Nojiri M~M and de~Roeck A 2007 {\em JHEP\/} {\bf 03} 046
  (\textit{Preprint} \eprint{hep-ph/0612060})

\bibitem{Feng:2015wqa}
Feng J~L, Iwamoto S, Shadmi Y and Tarem S 2015 {\em JHEP\/} {\bf 12} 166
  (\textit{Preprint} \eprint{1505.02996})

\bibitem{Jittoh:2005pq}
Jittoh T, Sato J, Shimomura T and Yamanaka M 2006 {\em Phys. Rev.\/} {\bf D73}
  055009 [Erratum: 2013 {\it Phys. Rev.} {\bf D87} 019901] (\textit{Preprint}
  \eprint{hep-ph/0512197})

\bibitem{Kaneko:2008re}
Kaneko S, Sato J, Shimomura T, Vives O and Yamanaka M 2013 {\em Phys. Rev.\/}
  {\bf D87} 039904 [Erratum: 2008 {\it Phys. Rev.} {\bf D78} 116013]
  (\textit{Preprint} \eprint{0811.0703})

\bibitem{Ellis:2003cw}
Ellis J~R, Olive K~A, Santoso Y and Spanos V~C 2003 {\em Phys. Lett.\/} {\bf
  B565} 176--182 (\textit{Preprint} \eprint{hep-ph/0303043})

\bibitem{ArkaniHamed:2004fb}
Arkani-Hamed N and Dimopoulos S 2005 {\em JHEP\/} {\bf 06} 073
  (\textit{Preprint} \eprint{hep-th/0405159})

\bibitem{ArkaniHamed:2004yi}
Arkani-Hamed N, Dimopoulos S, Giudice G~F and Romanino A 2005 {\em Nucl.
  Phys.\/} {\bf B709} 3--46 (\textit{Preprint} \eprint{hep-ph/0409232})

\bibitem{Evans:2012bf}
Evans J~A and Kats Y 2013 {\em JHEP\/} {\bf 04} 028 (\textit{Preprint}
  \eprint{1209.0764})

\bibitem{DiazCruz:2007fc}
Diaz-Cruz J~L, Ellis J~R, Olive K~A and Santoso Y 2007 {\em JHEP\/} {\bf 05}
  003 (\textit{Preprint} \eprint{hep-ph/0701229})

\bibitem{Giudice:1998xp}
Giudice G~F, Luty M~A, Murayama H and Rattazzi R 1998 {\em JHEP\/} {\bf 12} 027
  (\textit{Preprint} \eprint{hep-ph/9810442})

\bibitem{Randall:1998uk}
Randall L and Sundrum R 1999 {\em Nucl. Phys.\/} {\bf B557} 79--118
  (\textit{Preprint} \eprint{hep-th/9810155})

\bibitem{Aaboud:2019trc}
Aaboud M {\em et~al.\/} (ATLAS) 2019 {\em Phys. Rev.\/} {\bf D99} 092007
  (\textit{Preprint} \eprint{1902.01636})

\bibitem{Khachatryan:2016sfv}
Khachatryan V {\em et~al.\/} (CMS) 2016 {\em Phys. Rev.\/} {\bf D94} 112004
  (\textit{Preprint} \eprint{1609.08382})

\bibitem{ATLASsummary}
{ATLAS Collaboration} 2019 {Summary plots from the ATLAS Supersymmetry physics
  group}
  \url{https://atlas.web.cern.ch/Atlas/GROUPS/PHYSICS/CombinedSummaryPlots/SUSY/}

\bibitem{CMSsummary}
{CMS Collaboration} 2019 {CMS Exotica Summary plots for 13 TeV data}
  \url{https://twiki.cern.ch/twiki/bin/view/CMSPublic/SummaryPlotsEXO13TeV}

\bibitem{Acharya:2019vtb}
Acharya B {\em et~al.\/} (MoEDAL) 2019 {\em Phys. Rev. Lett.\/} {\bf 123}
  021802 (\textit{Preprint} \eprint{1903.08491})

\bibitem{Aad:2013gva}
Aad G {\em et~al.\/} (ATLAS) 2013 {\em Phys. Rev.\/} {\bf D88} 112003
  (\textit{Preprint} \eprint{1310.6584})

\bibitem{Sirunyan:2017sbs}
Sirunyan A~M {\em et~al.\/} (CMS) 2018 {\em JHEP\/} {\bf 05} 127
  (\textit{Preprint} \eprint{1801.00359})

\bibitem{Alwall:2014hca}
Alwall J, Frederix R, Frixione S, Hirschi V, Maltoni F, Mattelaer O, Shao H~S,
  Stelzer T, Torrielli P and Zaro M 2014 {\em JHEP\/} {\bf 07} 079
  (\textit{Preprint} \eprint{1405.0301})

\bibitem{Sjostrand:2007gs}
Sjostrand T, Mrenna S and Skands P~Z 2008 {\em Comput. Phys. Commun.\/} {\bf
  178} 852--867 (\textit{Preprint} \eprint{0710.3820})

\bibitem{Khachatryan:2015lla}
Khachatryan V {\em et~al.\/} (CMS) 2015 {\em Eur. Phys. J.\/} {\bf C75} 325
  (\textit{Preprint} \eprint{1502.02522})

\bibitem{Chatrchyan:2013oca}
Chatrchyan S {\em et~al.\/} (CMS) 2013 {\em JHEP\/} {\bf 07} 122
  (\textit{Preprint} \eprint{1305.0491})

\bibitem{Walkowiak:2018ief}
Walkowiak W (ATLAS) 2018 {\em PoS\/} {\bf BEAUTY2018} 055

\bibitem{LHCb:2018hiv}
{LHCb Collaboration} (LHCb) 2018 {Prospects for searches for long-lived
  particles after the LHCb detector upgrades} {LHCb-CONF-2018-006},
  CERN-LHCb-CONF-2018-006

\end{thebibliography}

\end{document}